\documentclass[fleqn]{2017SCGE}
\usepackage{multirow}
\usepackage{amsmath}

\makeatletter
\def\cpartlineleft#1{\@cpartlineleft#1\@nil}
\def\@cpartlineleft#1,#2\@nil{%
  \omit
  \@multicnt#1%
  \advance\@multispan\m@ne
  \ifnum\@multicnt=\@ne\@firstofone{&\omit}\fi
  \@multicnt#1%
  \advance\@multicnt-#1%
  \advance\@multispan\@ne
  \kern#2
        \leaders\hrule\@height\arrayrulewidth\hfill
        \leaders\hrule\@height\arrayrulewidth\hfill
  \cr
  \noalign{\vskip-\arrayrulewidth}}
\makeatother

\begin{document}

\ensubject{subject}

\ArticleType{Article}
\Year{2024}
\Month{XX}
\Vol{60}
\No{1}
\DOI{10.1007/s11432-016-0037-0}
\ArtNo{000000}
\ReceiveDate{XX X, 2024}
\AcceptDate{XX X, 2024}

\title{Dust processing in the terrestrial planet-forming region of the PDS\,70 disk}{Dust processing in the terrestrial planet-forming region of the PDS\,70 disk}

 
\author[1,2]{Yao Liu}{{yliu@swjtu.edu.cn}}
\author[2,3]{Dafa Li}{}
\author[2]{Hongchi Wang}{}
\author[2,3]{Haoran Feng}{}
\author[2]{Min Fang}{}
\author[2]{Fujun Du}{}
\author[4]{\\ Thomas Henning}{}
\author[5,4]{Giulia Perotti}{}

\AuthorMark{Liu}

\AuthorCitation{Liu et al.}

\address[1]{School of Physical Science and Technology, Southwest Jiaotong University, Chengdu 610031, China}
\address[2]{Purple Mountain Observatory, Chinese Academy of Sciences, 10 Yuanhua Road, Qixia District, Nanjing 210023, China}
\address[3]{School of Astronomy and Space Science, University of Science and Technology of China, 96 Jinzhai Road, Hefei 230026, China}
\address[4]{Max-Planck-Institut f\"ur Astronomie, K\"onigstuhl 17, D-69117 Heidelberg, Germany}
\address[5]{Niels Bohr Institute, University of Copenhagen, NBB BA2, Jagtvej 155A, 2200 Copenhagen, Denmark}

\abstract{Dust grains in protoplanetary disks are the building blocks of planets. Investigating the dust composition and size, and their variation over time, is crucial for understanding the planet formation process. The PDS\,70 disk is so far the only protoplanetary disk with concrete evidence for the presence of young planets. Mid-infrared spectra were obtained for PDS\,70 by the Infrared Spectrograph (IRS) on the Spitzer Space Telescope (SST) and the Mid-Infrared Instrument (MIRI) on the James Webb Space Telescope (JWST) in 2007 and 2022, respectively. In this work, we investigate the dust mineralogy through a detailed decomposition of the observed mid-infrared spectra. The results show that both the dust size and crystallinity increased by a factor of about two during the two epochs of observation, indicating evident dust processing in the terrestrial planet-forming region of the PDS\,70 disk. The dust size (${\sim}\,0.8\,\mu{\rm m}$) and crystallinity (${\sim}\,6\%$) in the PDS\,70 disk are similar to those of other disks, which implies that the two nascent planets, PDS\,70b and PDS\,70c located at radial distances of ${\sim}\,22\,{\rm AU}$ and ${\sim}\,34\,{\rm AU}$, do not have a significant impact on the dust processing in the inner disk. The flux densities at $\lambda\,{\gtrsim}\,16\,\mu{\rm m}$ measured by JWST/MIRI are only 60\% of those obtained by Spitzer/IRS. Based on self-consistent radiative transfer modeling, we found that such a strong variability in mid-infrared fluxes can be produced by adjustments to the dust density distribution and structure of the inner disk probably induced by planet-disk interaction.} 



\maketitle


\begin{multicols}{2}

\section{Introduction}
\label{sec:intro}

Protoplanetary disks are a natural product of the star formation process. Dust grains in these disks are the building blocks of terrestrial planets and the cores of giant planets \cite{Shu1987}. They gradually grow into pebbles through collisions and sticking \cite{1993ApJ...407..806C,1997ApJ...480..647D}, and then form planetesimals through mechanisms like streaming instability \cite{Youdin2005,Johansen2007}. Large dust grains settle towards the disk midplane due to the gravitational force exerted by the central star, thereby altering the geometric structure of the disk \cite{Dullemond2007}. Due to its large opacity, dust absorbs the stellar photons, and releases energy in the form of thermal re-emission, thereby establishing the temperature structure of the disk. Complex chemical reactions may occur on the surface of dust grains, and form organic molecules that are relevant to life \cite{Henning2013}. Therefore, investigating the dust properties is of great importance for understanding disk evolution and planet formation. In addition, comparing dust properties in protoplanetary disks with those in comets/meteorites also helps to clarify the origin of celestial bodies in our solar system.

\begin{figure*}[!t]
    \centering    
    \includegraphics[width=0.95\textwidth,angle=0]{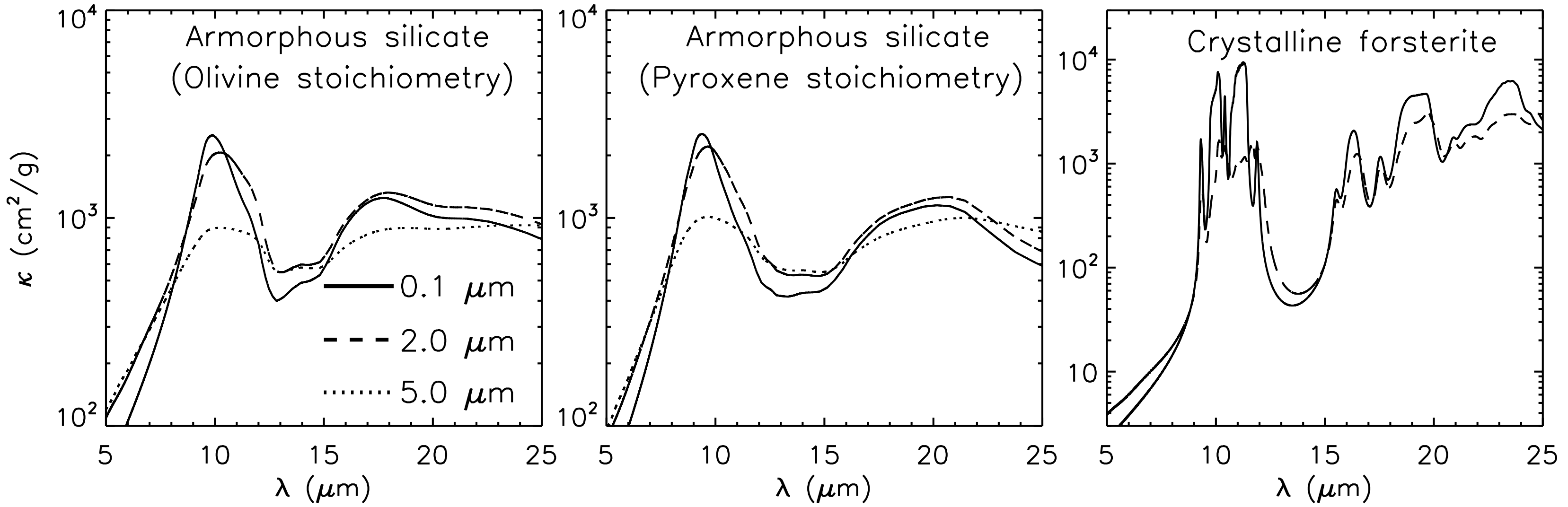}
    \caption{Mass absorption coefficients of amorphous silicates of olivine type (left panel) and pyroxene type (middle panel), and crystalline forsterite (right panel). The solid lines show the results for the grain size being $0.1\,\mu{\rm m}$. The dashed lines indicate the results for $2\,\mu{\rm m}$-sized dust grains. The dotted lines display the results for $5\,\mu{\rm m}$-sized dust grains. The results are calculated using the theory of distribution of hollow spheres \cite{Min2005}.}
    \label{fig:kappa}
\end{figure*}

Infrared spectra contain abundant information about the dust properties. Infrared spectroscopic observations indicate a large amount of silicate dust grains in protoplanetary disks \cite{Molster2002,Juhasz2010}. Silicates have strong emission features in the infrared, such as the $10\,\mu {\rm m}$ feature generated by the stretching vibrations of the silicon-oxygen bonds and the $18\,\mu {\rm m}$ feature attributed to the oxygen-silicon-oxygen bending vibration mode. For small (${\sim}\,0.1\,\mu {\rm m}$) dust particles, the emission features are relatively sharp. When the grain size is large (${\gtrsim}\,2\,\mu {\rm m}$), the emission features become flat and broad \cite{Boekel2005,Apai2005,Bouwman2008,Watson2009}, see the left panel of Figure~\ref{fig:kappa}. On the basis of the lattice structure, silicates can be divided into two categories: amorphous silicates and crystalline silicates. The amorphous silicates consist of silicon, oxygen and other atoms, which are arranged in a non-crystalline, disordered structure. Their $10\,\mu {\rm m}$ features are relatively broad and smooth. In high-temperature environments, amorphous silicates may undergo a heating and annealing process, and become crystallized \cite{Fabian2000}. Crystalline silicates have an ordered lattice structure, and their emission features exhibit narrow and sharp bands around $10\, \mu {\rm m}$, $16.3\,\mu{\rm m}$, $19.5\,\mu{\rm m}$ and $24.5\,\mu{\rm m}$, see the right panel of Figure~\ref{fig:kappa}. Therefore, analyzing the emission features present in infrared spectra is an important approach for studying dust size and mineralogy.

To extract dust properties from spectral decomposition, there are generally three steps. First, the mass absorption coefficients of dust with different components and sizes are calculated. Second, these mass absorption coefficients are incorporated into radiative transfer models of protoplanetary disks to calculate model spectra. Dust grains probed by infrared spectra mainly originate from the surface layer of the inner disk \cite{Jang2024a}, roughly corresponding to the terrestrial planet-forming region. The optical depth of the surface layer is relatively low, and the radiative transfer model can be approximated with either a single-temperature Planck function \cite{Boekel2005b} or a superposition of multiple-temperature Planck functions \cite{Juhasz2009}. Third, by combining dust grains with different compositions and sizes, and adjusting their mass fractions, the best match between model and observation can be achieved. The Markov Chain Monte Carlo algorithm \cite{Foreman-Mackey2013} and genetic algorithm \cite{Charbonneau1995} are often employed to conduct the parameter study.

PDS\,70 is a T Tauri star located in the Upper Centaurus-Lupus subgroup, part of the Scorpius-Centaurus association, at a distance of $d\,{=}\,112.4\,{\rm pc}$ \cite{Gaia2023}. It hosts a large protoplanetary disk, which is so far the only disk with concrete evidences for the presence of young planets. Keppler et al. \cite{Keppler2018} conducted high-resolution infrared imaging using the high-contrast imager Spectro-Polarimetric High-contrast Exoplanet REsearch (SPHERE) mounted at the Very Large Telescope (VLT), and discovered the protoplanet PDS\,70b that is located at a radial distance of 22\,AU from the central star. Haffert et al. \cite{Haffert2019} obtained H$\alpha$ emission maps using the Multi Unit Spectroscopic Explorer (MUSE) mounted at the VLT, and found that PDS\,70b is still accreting its surrounding material. Additionally, they identified another significant accretion feature at a radial location of 34\,AU, suggesting the presence of another planet, PDS\,70c. High-resolution millimeter images revealed a circumplanetary disk around PDS\,70c \cite{Isella2019,Benisty2021}, further supporting that PDS\,70c is a young planet. The Near Infrared Camera (NIRCam) is the primary imager of the James Webb Space Telescope (JWST). Recently, Christiaens et al. \cite{Christiaens2024} obtained NIRCam images of PDS 70 through the JWST MIRI mid-INfrared Disk Survey (MINDS, PID: 1282, PI: Thomas Henning \cite{Henning2024}). The authors identified a remarkable spiral-like signal connecting the outer disk with the position of PDS\,70c, and interpreted it as an accretion stream that feeds the circumplanetary disk of PDS\,70c. The extended data Figure\,1 of Perotti et al. \cite{Perotti2023} gives a sketch of the system. The interactions between PDS\,70b, PDS\,70c, and the natal disk have cleared material in the vicinity of the planet orbits, creating a prominent gap. The disk is thus separated into an inner disk from ${\sim}\,0.04\,\rm{AU}$ to $20\,\rm{AU}$ \cite{Keppler2018,Benisty2021} and an outer disk from ${\sim}\,60\,\rm{AU}$ to $120\,\rm{AU}$ \cite{Keppler2019}.

PDS\,70 was observed with the Infrared Spectrograph (IRS) installed on the Spitzer Space Telescope (SST) and the Mid-Infrared Instrument (MIRI) installed on the JWST in 2007 and 2022, respectively. In this work, we conduct a detailed spectral decomposition of the IRS and MIRI spectra, and investigate the dust properties in the terrestrial planet-forming region and their variation over a timespan of 15 years. Section~\ref{sec:obs} presents the observed IRS and MIRI spectra. Section~\ref{sec:specdecom} describes the approach used to decompose the spectra. We discuss the results in Section~\ref{sec:res}. In Section~\ref{sec:irsed}, we investigate the reasons why PDS\,70 shows a strong variability in the mid-infrared spectral energy distribution (SED). The paper ends up with a brief summary in Section~\ref{sec:sum}.

\section{Observation}
\label{sec:obs}

\begin{figure*}[!t]
    \centering
    \includegraphics[width=0.8\textwidth,angle=0]{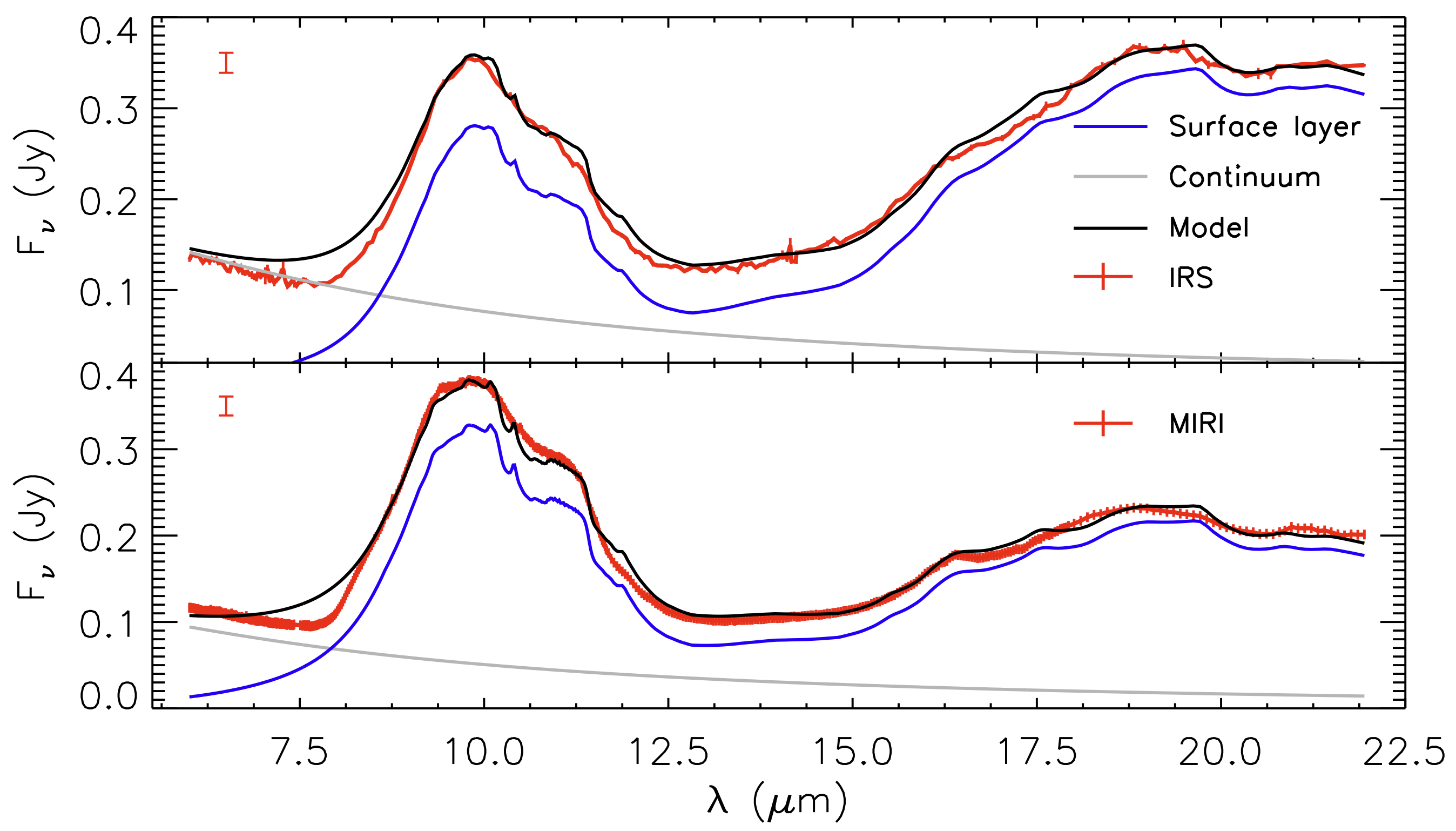}
\caption{Comparison between model spectra (black line) and observed spectra (red line) of PDS\,70. The upper panel shows the IRS spectrum, whereas the MIRI spectrum is shown in the bottom panel. For the model spectra, we also mark the contribution by the underlying continuum and dust emission from the disk surface layer, see Eq.~\ref{eq:newFnu}. The vertical bars on the IRS and MIRI spectra refer to the observational uncertainties. Note that the uncertainties of the IRS spectrum are barely seen. The vertical bar shown in the upper left corner of each panel indicates the finally-converged uncertainty obtained from the iterative fitting process, see Sect.~\ref{sec:fitapp}.}
\label{fig:fluxCom}
\end{figure*}

On July 30, 2007, mid-infrared spectra of PDS\,70 were obtained with the SST/IRS instrument (PID: 40679, PI: G. Rieke). The wavelength coverage is from 5.2\,$\mu\mathrm{m}$ to 35\,$\mu\mathrm{m}$, with a spectral resolution of $R\,{\approx}\,60{-}100$. PDS\,70 is one of the targets observed by the JWST/MIRI MINDS program. The MIRI spectra, obtained on August 1, 2022, cover a wavelength range from 4.9\,$\mu\mathrm{m}$ to 22.5\,$\mu\mathrm{m}$, and consist of four channels \cite{Argyriou2023}: channel 1 (4.9–7.65\,$\mu\mathrm{m}$, $R\,{\approx}\,3400$), channel 2 (7.51–11.71 $\mu\mathrm{m}$, $R\,{\approx}\,3000$), channel 3 (11.55–18.02\,$\mu\mathrm{m}$, $R\,{\approx}\,2400$), and channel 4 (17.71–22.5\,$\mu\mathrm{m}$, $R\,{\approx}\,1600$). In Perotti et al. \cite{Perotti2023}, the IRS and MIRI data were reduced using the recommended procedures from the Spitzer Science Center and the JWST Science Calibration Pipeline \cite{Christiaens2024}, respectively. The spectral resolution of MIRI is higher than that of the IRS, resulting in a denser wavelength sampling. Therefore, the original MIRI spectrum was rebinned by averaging 15 spectral points to further increase the signal-to-noise ratio. The red lines in Figure~\ref{fig:fluxCom} show the reduced spectra \footnote{\url{https://zenodo.org/records/7991022}} that are analyzed in this work.

Overall, the IRS and MIRI spectra exhibit similar silicate features at $10\,\mu{\rm m}$ and $19\,\mu{\rm m}$. At wavelengths $\lambda\,{\lesssim}\,16\,\mu{\rm m}$, PDS\,70 did not show a significant variation in the flux density. However, at wavelengths $\lambda\,{\gtrsim}\,16\,\mu{\rm m}$, the flux densities measured by MIRI are only about 60\% of the observed IRS levels. Observations at infrared wavelengths trace the warm inner disk. Therefore, the mid-infrared variability suggests that dust properties or the structure of the inner disk may change during the two epochs of observation \cite{Flaherty2012,Flaherty2016}.

\section{Spectral decomposition}
\label{sec:specdecom}

\subsection{Spectral model}
\label{sec:fitmod}

To constrain the dust properties, we conducted a detailed analysis of the IRS and MIRI spectra in the wavelength range from 6\,$\mu\mathrm{m}$ to 22\,$\mu\mathrm{m}$. This wide wavelength range of dust emission is expected to originate from a range of radius in the disk that have a radial temperature distribution. Therefore, we prescribed the model spectrum to be the sum of the dust emission feature and the underlying continuum, both of which are characterized by a power-law temperature profile. Specifically, the model spectra are expressed as 
\begin{equation}
\begin{aligned}
\label{eq:Fnu}
F_\nu\,{=}\,& \sum_{i=1}^N \sum_{j=1}^M C_{i, j} \kappa_{i, j} \int_{{R_{\mathrm{sur.in}}}}^{R_\mathrm{sur.out}} \frac{2 \pi r}{d^2} B_\nu\left(T_{\mathrm{sur}}(r)\right) \mathrm{d} r \\
& + C_0 \int_{R_{\mathrm {cont.in} }}^{R_{\mathrm {cont.out}}} \frac{2 \pi r}{d^2} B_\nu\left(T_{\mathrm {cont}}(r)\right) \mathrm{d} r,
\end{aligned}
\end{equation}
where $B_{\nu}$ refers to the Planck function at the given frequency $\nu$. The first term in Equation~\ref{eq:Fnu} represents the dust emission feature, in which $N$ and $M$ are the number of dust components and grain sizes, respectively. The absorption coefficient for the $i$-th dust component and the $j$-th grain size is denoted as $\kappa_{i, j}$. The mass fractions of different components and sizes of dust are calculated through $m_{i, j}\,{=}\,C_{i, j}/\sum_{i=1}^N \sum_{j=1}^M C_{i, j}$.

The temperatures responsible for the dust emission feature and the continuum are depicted as $T_{\mathrm {sur}}(r)$ and $T_{\mathrm {cont}}(r)$, respectively, both of which follow a power-law distribution:
\begin{equation}
\label{eq:powert}
\begin{aligned}
T_{\mathrm {sur}}(r) & =T_{\mathrm {sur.max}}\left(\frac{r}{R_{\mathrm {sur.in}}}\right)^{q_{\mathrm {sur}}}, \\
T_{\mathrm {cont}}(r) & =T_{\mathrm {cont.max }}\left(\frac{r}{R_{\mathrm {cont.in}}}\right)^{q_{\mathrm {cont}}}.
\end{aligned}
\end{equation}

Following Juh{\'a}sz et al. \cite{Juhasz2009}, the integrals over radius $r$ in Equation~\ref{eq:Fnu} can be rewritten as integrals over temperature $T$: 
\begin{equation}
\label{eq:newFnu}
\begin{aligned}
F_\nu\,{=}\,& \sum_{i=1}^N \sum_{j=1}^M D_{i, j} \kappa_{i, j} \int_{T_{\mathrm {sur.max }}}^{T_{\mathrm {sur.min }}} B_\nu(T) T^{\frac{2-q_{\mathrm {sur}}}{q_{\mathrm {sur}}}} \mathrm{d} T \\
 & + D_0 \int_{T_{\mathrm{cont.max}}}^{T_{\mathrm{cont.min}}} B_\nu(T) T^{\frac{2-q_{\mathrm{cont}}}{q_{\mathrm{cont}}}} \mathrm{d} T \,,
\end{aligned}
\end{equation}
where
\begin{align}
D_{i, j}&=\frac{2\pi T_{\mathrm{sur.max}}^{-2 / q_{\mathrm{sur}}} R_{\mathrm{sur.in}}^2}{d^2 q_{\mathrm{sur}}} C_{i, j}\\
D_0&=\frac{2\pi T_{\mathrm{cont.max}}^{-2 / q_{\mathrm{cont}}} R_{\mathrm{cont.in}}^2}{d^2 q_{\mathrm{cont}}} C_0.
\end{align}
In the fitting process, $T_{\mathrm {cont.max}}$, $T_{\mathrm {cont.min}}$, $q_{\mathrm {cont}}$, $T_{\mathrm {sur.max}}$, $T_{\mathrm {sur.min}}$, $q_{\mathrm {sur}}$, $D_{i,j}$ and $D_0$ are free parameters. Finally, the mass fraction of each dust component and each grain size is calculated via $m_{i, j}=D_{i, j} / \sum_{i=1}^N \sum_{j=1}^M D_{i, j}$.

We considered three different dust compositions including amorphous silicates with olivine stoichiometry \cite{Jager2003}, amorphous silicates with pyroxene stoichiometry \cite{Dorschner1995} and crystalline forsterite \cite{Sogawa2006}. Previous studies have shown that these compositions are necessary for reproducing the observed emission features of protoplanetary disks \cite{Bouwman2008,Juhasz2010}. For amorphous silicates with olivine and pyroxene stoichiometry, we considered three grain sizes, i.e., 0.1\,$\mu\mathrm{m}$, 2\,$\mu\mathrm{m}$ and 5\,$\mu\mathrm{m}$. For crystalline forsterite, we only considered two grain sizes: 0.1\,$\mu\mathrm{m}$ and 2\,$\mu\mathrm{m}$. Based on the theory of the distribution of hollow spheres (DHS) \cite{Min2005}, we calculated the mass absorption coefficients using the complex refractive indices of each type of dust. The results are shown in Figure~\ref{fig:kappa}.

\subsection{Fitting approach}
\label{sec:fitapp}

We used the MultiNest Bayesian fitting algorithm and the \texttt{PyMultiNest} package to search for the best-fit parameter set \cite{Feroz2008,Buchner2014}. In \texttt{PyMultiNest}, the number of live points is a key input that influences the sampling efficiency and fitting accuracy. Taking a large number increases the computational cost. To balance the fitting accuracy and computation time, we set the number of live points to 2,000 that is enough for searching for optimal solutions in the parameter space of our work, since using a larger value obtains similar results. We note that the choice for the number of live points is not universal, and it depends on the complexity of the model and the dimensionality of the parameter space. 

In the fitting process, the uncertainty of the observed spectra was initially set to $\sigma_{\nu}\,{=}\,0.01\,{\rm Jy}$, constant over the wavelength. Then, we fit the spectra, and derived the standard deviation of the residuals between the best-fit model and the observed spectra. The standard deviation was taken as the uncertainty of the observed spectra for the next round of fitting. Such a process was iterated until the standard deviation of the residual spectrum converges. Typically, it needs four iterations to reach the convergence.  We devised this kind of iterative fitting approach to account for both the model uncertainty and spectral noise. The finally-converged uncertainty is ${\sim}\,0.01\,\rm{Jy}$ for both spectra, which is indicated with the vertical bar in the upper left corner of each panel in Figure~\ref{fig:fluxCom}. Literature works introduced different strategies to deal with the uncertainties. For instance, to fit the MIRI spectrum, Kaeufer et al. \cite{Kaeufer2024b} defined $\sigma_{\nu}$ to be proportional to the observed flux $\sigma_{\nu}\,{=}\,\xi\,F_{\nu.{\rm obs}}$, where $\xi$ is a free parameter. Both their result and our experiment suggest that the absolute values of the observational uncertainties do not have a significant effect on the retrieved dust properties as long as the signal-to-noise ratio of the spectrum is high.

\subsection{Results}
\label{sec:res}

In Figure~\ref{fig:fluxCom}, the black lines represent the best fits, showing a good match between the model spectra and observations. The best-fit parameter values together with their uncertainties are summarized in Table~\ref{tab:specparas}, whereas Table~\ref{tab:comp} lists the mass percentages of different dust components. We calculated the mass-averaged dust size using these mass percentages as weighting factors. The averaged dust size obtained from fitting the IRS spectrum is $a^{\mathrm{IRS}}_{\mathrm{avg}} = 0.33 ^{+0.13}_{-0.09}\,\mu\mathrm{m}$, whereas fitting the MIRI spectrum results in an averaged size of $a^{\mathrm{MIRI}}_{\mathrm{avg}} = 0.83 ^{+0.12}_{-0.14}\,\mu\mathrm{m}$. Comparing the results of the two fits, there is a two times of increment in dust size.  Birnstiel et al. \cite{2012A&A...539A.148B} pointed out that the timescale for dust growth, doubling in grain size, is approximately $\tau_{\mathrm {grow }}\,{\simeq}\,\frac{1}{\Omega_{\mathrm{k}} \epsilon}$. In the formula, $\epsilon$ is the dust-to-gas mass ratio, and $\Omega_{\mathrm{k}}\,{=}\,\sqrt{GM_{\star}/r^3}$ is the Keplerian angular velocity, where $G$ is the gravitational constant, $M_{\star}\,{=}\,0.8\,M_{\odot}$ is the mass of the central star \cite{Muller2018}. At $r\,{=}\,5\,\rm{AU}$, we took $\epsilon$ to be 0.1 \cite{Portilla2023}, so the required time for dust grains growing up by a factor of two is ${\sim}\,20$ years that is close to the timespan between the IRS and MIRI observations, i.e., 15 years.

We derived the dust crystallinity by summing up the mass percentages of the crystalline forsterite. Fitting the IRS spectrum yields a crystallinity of 3.4\%. The MIRI spectrum reflects an increased crystallinity of 6.3\%. There are different mechanisms proposed for the formation of crystals. For instance, high temperature environments can anneal amorphous silicate grains into crystalline grains. Moreover, high temperature gas phase condensation of silicates can also form crystals. These conditions are likely satisfied in the inner disk since it is exposed to the stellar radiation or outbursts \cite{Abraham2009}, and it also can be impacted by shock waves \cite{Harker2002}. More observations, for example monitoring campaigns in the optical, will help to identify which mechanism is responsible for the ongoing formation of crystals in the PDS\,70 disk. Both the mass-averaged grain size and crystallinity increased by a factor of about two over a period of 15 years, suggesting that dust grains in the terrestrial planet-forming region of the PDS\,70 disk have undergone evident processing. Dust evolution in protoplanetary disks is governed by the collisional growth \cite{Birnstiel2024}, and is also influenced by the mechanisms for disk dispersal, such as accretion process and disk winds \cite{Alexander2014}. The larger dust grains revealed by the MIRI data is likely produced via grain growth within the inner disk, since the timescale for grain growth (${\sim}\,20$ years) is close to the timespan between the IRS and MIRI observations. High-resolution (sub-)millimeter observations suggest that large dust grains are present in the outer ring of the PDS\,70 disk \cite{Doi2024,LiuH2024}. Material in the outer ring may also permeate through the gap region, and continuously feed the inner disk \cite{Pinilla2024,Petrovic2024}. Our analysis is performed for infrared spectra of only two separate epochs. More spectroscopic observations at infrared wavelengths are required to monitor the changes of dust size and crystallinity, and further investigate the dust evolution in the terrestrial planet-forming zone of the PDS\,70 disk.

\begin{table}[H]
  \centering
    \caption{Best-fit parameter values of the spectral models.}
	\linespread{1.1}\selectfont
	\setlength{\tabcolsep}{6pt}
    \begin{tabular}{lcc}
    \hline
    Parameter    &    IRS spectrum    &  MIRI spectrum \\
	\hline
    $T_{\rm sur.max}$ [K]     &   $557.1^{+64.17}_{-40.75}$   &  $1029.0^{+122.4}_{-87.6}$  \\
    $T_{\rm sur.min}$ [K]     &   $102.5^{+3.42}_{-1.77}$     &  $110.3^{+15.52}_{-7.60}$   \\
    $q_{\rm sur}$             &   $-0.52^{+0.02}_{-0.02}$     &  $-0.72^{+0.02}_{-0.02}$    \\
    $T_{\rm cont.max}$ [K]    &   $1483.0^{+11.71}_{-21.06}$  &  $1492.0^{+5.73}_{-10.8}$   \\
    $T_{\rm cont.min}$ [K]    &   $1404.0^{+32.92}_{-48.51}$  &  $1407.0^{+27.42}_{-37.78}$ \\
    $q_{\rm cont}$            &   $-1.08^{+0.08}_{-0.09}$     &  $-1.94^{+0.08}_{-0.05}$    \\
	$D_{0}$                   &   $0.61^{+0.25}_{-0.27}$      &  $0.84^{+0.11}_{-0.19}$     \\
    \hline
    \label{tab:specparas}
    \end{tabular}
\end{table}

\begin{table*}[!t]
  \centering
    \caption{Best-fit mass percentages of different dust components}
    \resizebox{0.9\textwidth}{!}{\begin{tabular}{lcccccccc}
    \toprule
	\multirow{2}{*}{Dust species}  & \multirow{2}{*}{Chemical form} & \multicolumn{3}{c}{IRS spectrum} & & \multicolumn{3}{c}{MIRI spectrum} \\
		    \cline{3-5} \cline{7-9}
            &  &  0.1\,$\mu\mathrm{m}$  & 2\,$\mu\mathrm{m}$ &  5\,$\mu\mathrm{m}$  &  & 0.1\,$\mu\mathrm{m}$  &  2\,$\mu\mathrm{m}$  &  5\,$\mu\mathrm{m}$  \\
    \midrule
    Amorphous silicate   & \multirow{2}{*}{$\rm{Mg_2SiO_4}$} &  \multirow{2}{*}{$48.79^{+4.68}_{-6.70}$} & \multirow{2}{*}{$0.70^{+0.92}_{-0.50}$} & \multirow{2}{*}{$1.05^{+1.27}_{-0.75}$} & & \multirow{2}{*}{$14.35^{+2.68}_{-3.06}$} & \multirow{2}{*}{$32.29^{+5.00}_{-6.68}$} & \multirow{2}{*}{$0.23^{+0.35}_{-0.17}$}  \\
	(Olivine stoichiometry) \\
    Amorphous silicate   & \multirow{2}{*}{$\rm{MgSiO_3}$}   &  \multirow{2}{*}{$39.90^{+5.65}_{-6.31}$} & \multirow{2}{*}{$4.83^{+3.60}_{-2.97}$} & \multirow{2}{*}{$1.34^{+1.57}_{-0.93}$} &  & \multirow{2}{*}{$44.81^{+5.25}_{-8.67}$} & \multirow{2}{*}{$1.84^{+2.63}_{-1.36}$} & \multirow{2}{*}{$0.23^{+0.35}_{-0.17}$} \\
	(Pyroxene stoichiometry) \\
    Crystalline forsterite & $\rm{Mg_2SiO_4}$ & $2.78^{+0.38}_{-0.45}$ & $0.62^{+0.49}_{-0.39}$ & $-$ &  & $3.20^{+0.44}_{-0.62}$ & $3.05^{+0.98}_{-0.94}$ & $-$ \\
    \midrule
    Crystallinity & & & $3.40^{+0.62}_{-0.60}$ & & & & $6.25^{+1.08}_{-1.13}$ & \\
    Averaged dust size   [$\mu$m] & & & $0.33^{+0.13}_{-0.09}$ & & & & $0.83^{+0.12}_{-0.14}$ & \\
    \bottomrule
    \label{tab:comp}
    \end{tabular}}
\end{table*}

\begin{figure*}[!t]
\centering \includegraphics[width=0.8\textwidth,angle=0]{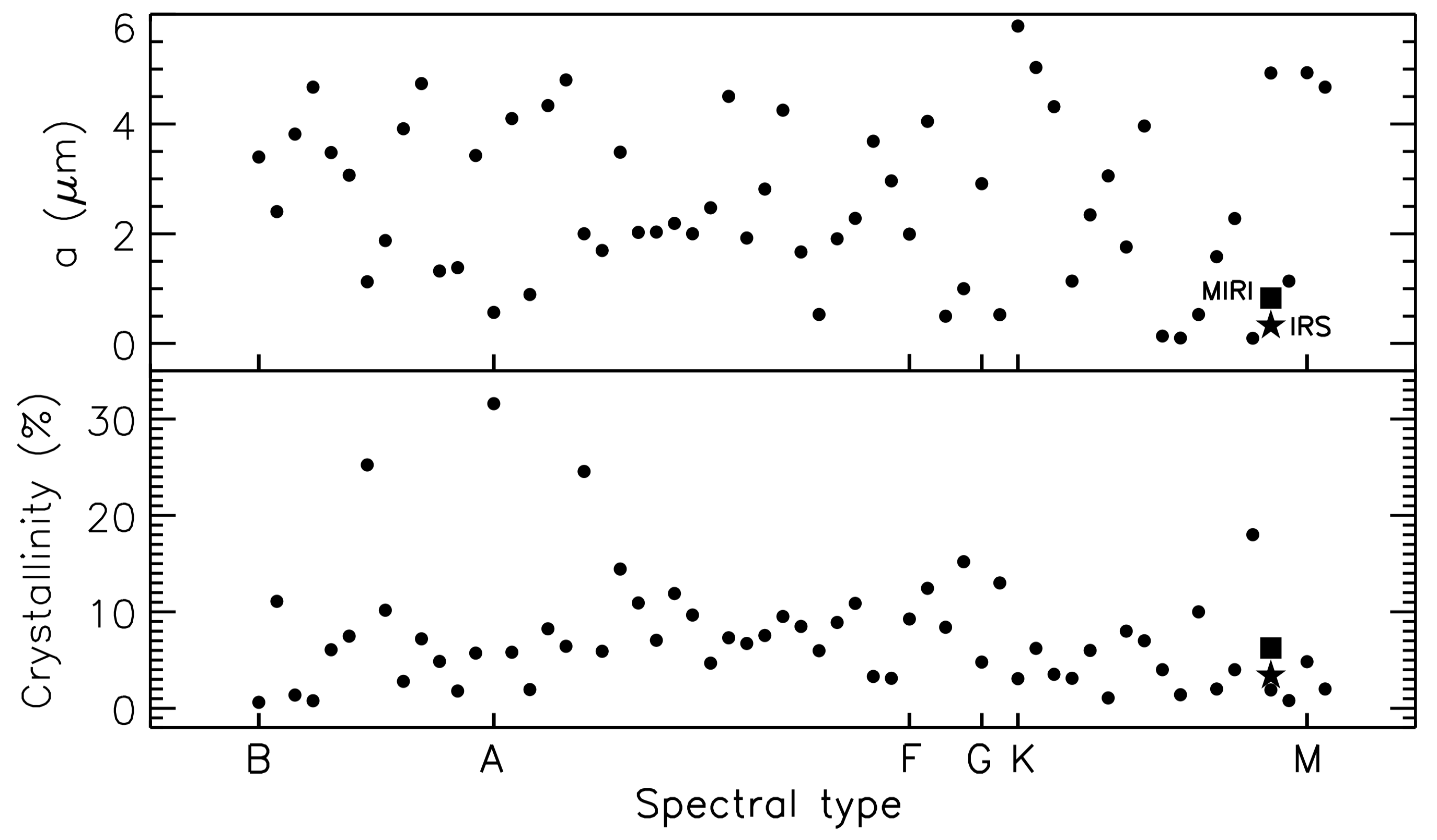}
\caption{Dust grain size (upper panel) and crystallinity (bottom panel) in protoplanetary disks as a function of spectral type. The black star and black square represent the results obtained by fitting the IRS and MIRI spectra of PDS\,70, respectively, while the black dots refer to the results of other disks analyzed by Sicilia-Aguilar et al. \cite{Sicilia-Aguilar2007}, Bouwman et al. \cite{Bouwman2008} and Juh{\'a}sz et al. \cite{Juhasz2010}.} 
\label{fig:other}
\end{figure*}

We also compared the dust properties of PDS\,70 with those of other protoplanetary disks. Figure~\ref{fig:other} displays the averaged size and crystallinity of dust grains in a sample of protoplanetary disks, along with the spectral type of their host central stars. The results obtained from fitting the IRS and MIRI spectra of PDS\,70 are indicated with the black star and black square, respectively. The black dots show the results presented by Sicilia-Aguilar et al. \cite{Sicilia-Aguilar2007}, Bouwman et al. \cite{Bouwman2008} and Juh{\'a}sz et al. \cite{Juhasz2010}, who analyzed infrared spectra of 59 protoplanetary disks using similar methods for the spectral decomposition. As can be seen, the range of dust size is relatively large, and PDS\,70 falls toward the lower end of the distribution. The crystallinity of most targets is below ${\sim}\,10\%$, with only few disks reaching $20\,{-}\,30\%$. For PDS\,70, the crystallinity shown with the IRS spectrum is $3.4\%$, while it is $6.3\%$ revealed by the MIRI data (see Table~\ref{tab:comp}), both of which are similar to the values of other disks. To the best of our knowledge, there is currently no strong evidence for the presence of (proto-)planets in any of the other protoplanetary disks. Therefore, our comparison suggests that although PDS\,70 hosts two young giant planets, dust properties in the terrestrial planet-forming region do not exhibit particular characteristics. Furthermore, the statistical analysis also shows that there is no strong correlation between the mass-averaged dust size, crystallinity and the spectral type of the host central star.

\subsection{Carvets of the model}
Figure~\ref{fig:fluxCom} shows that the central wavelength of the ${\sim}\,19.5\,\mu{\rm m}$ dust feature is not well reproduced. This suggests that the real dust composition in the PDS\,70 disk may be more complex than assumed in this work. Using a different grain shape model, for instance Gaussian random field particles \cite{Min2007}, may improve the quality of fit as well \cite{Jang2024b}. The discrepancies between model and observation around ${\sim}\,8\,\mu{\rm m}$ can be mitigated by considering more grain sizes for each dust species or introducing more components to describe the underlying continuum in the fitting procedure \cite{Kaeufer2024a}. However, implementing the modifications would increase the number of free parameters, and a Bayesian comparison may reject the new model despite a better fit to the spectrum at these local wavelengths. In addition, the discrepancies may come from the optical constants which never be perfect for the dust material in protoplanetary disks. 

\section{Variability in the mid-infrared SED}
\label{sec:irsed}

Through a detailed spectral decomposition, we have constrained the dust properties by capturing the width, sharpness, and central wavelength of distinctive dust features, e.g., at ${\sim}\,9.8\,\mu{\rm m}$, ${\sim}\,11.3\,\mu{\rm m}$, ${\sim}\,16.3\,\mu{\rm m}$ and ${\sim}\,19.5\,\mu{\rm m}$. As described by Eq.~\ref{eq:powert}, the power-law temperature profiles responsible for the dust features and the underlying continuum are freely explored in the parameter study. Consequently, with this type of analysis, it is difficult to deliver key information about disk structure and dust density distribution that determines the absolute level of flux densities in a self-consistent way. Figure~\ref{fig:fluxCom} shows that flux densities at wavelengths $\lambda\,{\gtrsim}\,16\,\mu\mathrm{m}$ in the MIRI spectrum are only 60\% of the IRS measurement. Such a prominent mid-infrared variability cannot be explained by the flux calibration uncertainty alone \cite{Argyriou2023}. Instead, it is likely caused by changes in stellar heating, inner disk structure, and dust density distribution during the two epochs of observation. In this section, we perform radiative transfer simulations to explain the variation in the absolute level of mid-infrared flux densities using more realistic disk models. The dust properties are a necessary input for radiative transfer simulations, and can be explored alongside with the disk structure in parameter studies \cite{Woitke2019}. To reduce the degrees of freedom, we directly incorporated the dust composition derived from the spectral decomposition into the radiative transfer models. This strategy has been demonstrated to be successful in interpreting infrared spectra of protoplanetary disks \cite{Grimble2024}.

\subsection{Theoretical speculations}
\label{sec:theory}

Changes in stellar radiation can alter the temperature and structure of the inner disk, thereby affecting the flux in the mid-infrared \cite{Muzerolle2009,espaillat2011}. Variations in stellar heating would cause an overall increase or decrease in infrared flux. However, in the case of PDS\,70, a significant flux reduction is observed only at wavelengths of $\lambda\,{\gtrsim}\,16\,\mu\mathrm{m}$, and there is actually a mild increment in flux at $\lambda\,{\sim}\,10\,\mu\mathrm{m}$, see Figure~\ref{fig:fluxCom}. Monitoring observations by the Wide-Field Infrared Survey Explorer (WISE) confirm that there are no significant changes in the W1 ($\lambda\,{=}\,3.4\,\mu\mathrm{m}$) and W2 ($\lambda\,{=}\,4.6\,\mu\mathrm{m}$) magnitudes, see the extended data Figure 6 of Perotti et al. \cite{Perotti2023}. Therefore, pure stellar radiation cannot explain the peculiar infrared variability of PDS\,70.

For T Tauri stars, dust temperature in the inner disk ranges from ${\sim}\,1000\,{\rm K}$ to ${\sim}\,100\,{\rm K}$ \cite{Chiang1997}, see Panel (b) of Figure~\ref{fig:moredust}. Hence, mid-infrared continuum fluxes mainly originate from the inner disk \cite{Chiang1997,Dullemond2001}, and the absolute emission level is determined by the temperature structure and dust density distribution. The scale height and flaring index characterize the vertical geometry of the disk, and have a significant impact on the heating of disks. Therefore, the variation of mid-infrared SED might be attributed to adjustments of scale height and/or flaring index of the inner disk. Moreover, the interaction between PDS\,70b and PDS\,70c with the natal disk has created a prominent gap, preventing a stable replenishment of material from the outer disk to the inner disk. Consequently, the inner disk might lose material gradually due to disk wind and accretion process, and therefore the infrared flux decreases accordingly. From a thorough analysis of VLT/SPHERE images, Mesa et al. \cite{Mesa2019} proposed a third protoplanet candidate at a short separation of $r\,{\sim}\,13\,\rm{AU}$ from the central star, close to the outer edge of the inner disk. The third protoplanet candidate was also found near the brightest signals in the JWST/NIRCam image presented by Christiaens et al. \cite{Christiaens2024}. The interaction between the inner disk and the third protoplanet candidate can also cause a redistribution of dust grains within the inner disk, which in turn alters the mid-infrared SED. These mechanisms might work together in PDS\,70 to explain the observed fact that the ${\sim}\,10\,\mu\mathrm{m}$ silicate feature becomes a bit stronger and flux densities at $\lambda\,{\gtrsim}\,16\,\mu\mathrm{m}$ get obviously reduced over a timespan of 15 years. 

\subsection{Radiative transfer modeling}

To test the above hypothesis, we ran radiative transfer models using the \texttt{RADMC-3D} \footnote{\url{https://www.ita.uni-heidelberg.de/~dullemond/software/radmc-3d/}} code \cite{Dullemond2012}. We emphasize that the motivation of our modeling is not to pursue perfect fits to the IRS and MIRI spectra, but to investigate how the inner disk structure affects the infrared SED in a self-consistent way. Changes in stellar heating was not considered. 

\begin{figure*}[!t]
\centering \includegraphics[width=0.95\textwidth,angle=0]{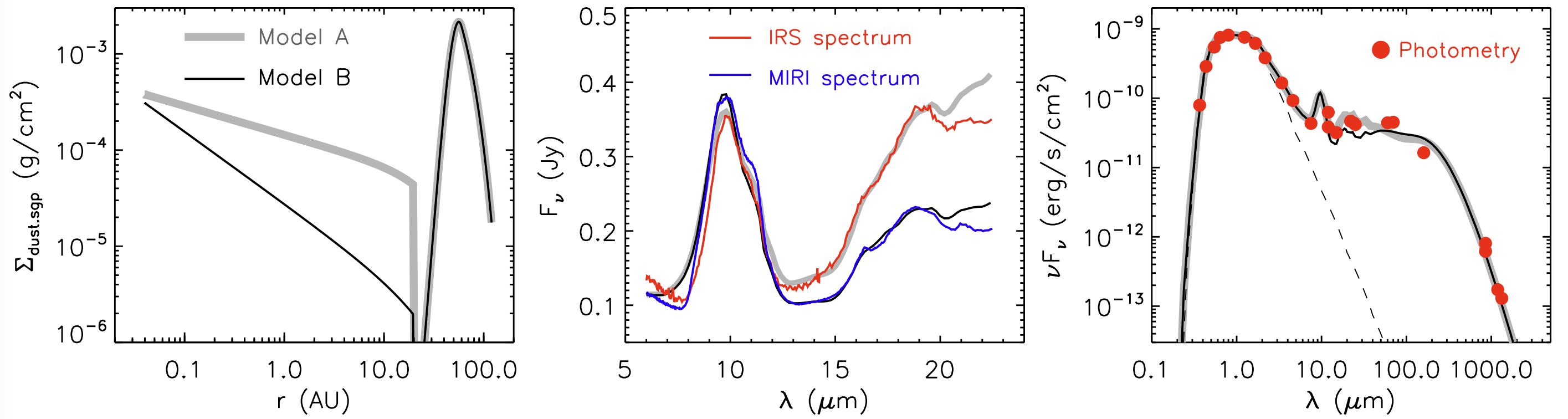}
\caption{Radiative transfer models of PDS\,70. {\it Left panel: } dust surface densities of model\,A (grey line) and model\,B (black line). Model parameters are given in Table~\ref{tab:paras}. Note that the surface densities of the outer disk for both models are identical. {\it Middle panel: } comparison of infrared spectrum between model and observation. The IRS data is indicated with the red line, whereas the blue line represents the MIRI spectrum. {\it Right panel: } comparison between the model SED and photometry from  optical to millimeter wavelengths. Data points are collected from Gregorio-Hetem et al. \cite{Gregorio-Hetem1992}, Hashimoto et al. \cite{Hashimoto2012}, Benisty et al. \cite{Benisty2021}, and Facchini et al. \cite{Facchini2021}.}
\label{fig:surdens}
\end{figure*}

\begin{table*}[!t]
  \centering
    \caption{Parameter values of the radiative transfer models shown in Figure~\ref{fig:surdens}.}
	\linespread{1.0}\selectfont
	\setlength{\tabcolsep}{6pt}
    \begin{tabular}{lcccc}
    \hline
	 \multirow{2}{*}{Parameter}   &    \multicolumn{2}{c}{Model\,A}    &    \multicolumn{2}{c}{Model\,B}  \\
	                                 \cline{2-3}         \cpartlineleft{4,0.8em}\cline{5-5}
	                              &    Inner disk    &  Outer disk     &     Inner disk   &  Outer disk   \\						     
    \hline
	 $r_{\rm in}$  [AU]           &    0.04          &      60             &    0.04          &     60        \\ 
	 $r_{\rm out}$ [AU]           &    20            &      120            &    20            &     120       \\
     $\beta$                      &    1.1           &      1.1            &    1.1           &     1.1       \\
	 $H_{\rm 100.LGP}$ [AU]       &    2             &      2              &    2             &     2         \\
	 $H_{\rm 100.SGP}$ [AU]       &    8             &      8              &    12            &     8        \\
	 $p$                          &    0.3           &      0.3            &    0.7           &     0.3       \\
     $M_{\rm dust}$ [$M_{\odot}$] & $8.4\times10^{-9}$ & $8\times10^{-5}$  & $5\times10^{-10}$  & $8\times10^{-5}$  \\
     $\eta$                       & $3\,{\times}\,10^{-3}$    &    1    &  $5\,{\times}\,10^{-4}$  & 1 \\
    \hline
    \label{tab:paras}
    \end{tabular}
\end{table*}

\subsubsection{Model setup}
\label{sec:rtmodel}

Keppler et al. \cite{Keppler2018} built a radiative transfer model that well reproduces the near-infrared polarimetric images and broadband SED of PDS\,70. Our model is constructed based on their setup. We assumed that the disk is passively heated by stellar irradiation, which is characterized by an effective temperature of $T_{\rm eff}\,{=}\,4500\,{\rm K}$ and a luminosity of $L_{\star}\,{=}\,0.45\,L_{\odot}$ \cite{Gregorio-Hetem2002,Dong2012}. The disk has two distinct dust grain populations, i.e., a small grain population (SGP) and a large grain population (LGP). The LGP dominates the total dust mass, and it is distributed only in the outer disk. We fixed its mass fraction to $f_{\rm LGP}\,{=}\,97\%$. The disk scale height follows a power law 
\begin{equation}
H\,{=}\,H_{100}\left(\frac{r}{100\,\rm{AU}}\right)^{\beta}, 
\end{equation}
where $r$ is the distance from the central star measured in the disk midplane, $\beta$ is the flaring index, and $H_{100}$ refers to the scale height at a characteristic radius of $r\,{=}\,100\,\rm{AU}$. For the LGP, we took a small value of $H_{\rm 100.LGP}\,{=}\,2\,\rm{AU}$, mimicking that large dust grains are concentrated close to the disk midplane due to the effect of dust settling \cite{Dubrulle1995,Dullemond2004,DAlessio2006}. We left $H_{\rm 100.SGP}$ and $\beta$ as free parameters, because they have an important impact on the dust temperature and therefore the spectral profile. The dust volume density is parameterized as 
\begin{equation}
\rho_{\rm{SGP}}(r,z)\,{=}\,\frac{(1-f_{\rm LGP})\,\Sigma_{\rm dust}}{\sqrt{2\pi}\,H_{\rm SGP}}\,\exp\left[-\frac{1}{2}\left(\frac{z}{H_{\rm SGP}}\right)^2\right], \\
\label{eqn:sgp}
\end{equation}
\begin{equation}
\rho_{\rm{LGP}}(r,z)\,{=}\,\frac{f_{\rm LGP}\,\Sigma_{\rm dust}}{\sqrt{2\pi}\,H_{\rm LGP}}\,\exp\left[-\frac{1}{2}\left(\frac{z}{H_{\rm LGP}}\right)^2\right]. \\
\label{eqn:lgp}
\end{equation}
The dust surface density $\Sigma_{\rm dust}$ follows a power-law profile plus an exponential taper
\begin{equation}
\Sigma_{\rm dust}(r)\,{=}\,\Sigma_{0}\left(\frac{r}{R_{\rm c}}\right)^{-p}{\rm exp}\left[-\left(\frac{r}{R_{\rm c}}\right)^{2-p}\right], \\
\label{eqn:sigma}
\end{equation}
where the power-law index $p$ quantifies the steepness of the profile, and $\Sigma_{0}$ is the proportionality constant determined by normalizing the total dust mass ($M_{\rm dust}$) in the entire disk. The characteristic radius $R_{\rm c}$ was fixed to 40\,AU. We multiplied $\Sigma_{\rm dust}$ between the outer radius of the inner disk and the inner radius of the outer disk by a factor of $10^{-10}$ in order to introduce a deep gap. To ensure a smooth transition from the gap region to the outer disk, we chose an outer disk radius of 60\,AU, inward of which $\Sigma_{\rm dust}$ is multiplied by a Gaussian profile with a standard deviation of 8\,AU. To fit the mid-infrared spectra, we found that $\Sigma_{\rm dust}$ in the inner disk have to be depleted with respect to that of the outer disk by a factor of $\eta$. Such a depletion of material in the inner disk is also required to reproduce the Subaru H-band polarized light image \cite{Dong2012}. The grey solid line in the left panel of Figure~\ref{fig:surdens} shows the general shape of $\Sigma_{\rm dust}$, and it was found to reproduce the IRS spectrum well, see Sect.~\ref{sec:fitirs}. 

Dust opacities of pure silicate grains are low in the optical and near-infrared wavelength domains, yielding a low capacity of absorbing stellar photons. The resulting dust temperature is not sufficiently high to produce strong silicate features in self-consistent radiative transfer models, see \ref{sec:moredust} for a detailed discussion. Therefore, as a first step, we separately mixed the refractive indices of amorphous silicates with olivine stoichiometry, amorphous silicates with pyroxene stoichiometry and crystalline forsterite with those of amorphous carbon grains \cite{Zubko1996}. We used the Bruggeman mixing rule and fixed the volume fraction of carbon to 10\%. Then, we calculated mass absorption and scattering coefficients of each of the mixture with the DHS theory. In the calculation, we assumed that the grain size distribution follows a power law ${\rm d}n(a)\,{\propto}\,{a^{-3.5}} {\rm d}a$. For the SGP, the minimum grain size ($a_{\rm{min}}$) and maximum grain size ($a_{\rm{max}}$) were fixed to $0.005\,\mu{\rm m}$ and $0.1\,\mu{\rm m}$, respectively. The power-law index and minimum grain size of the distribution were assumed to be identical to those found for the interstellar medium dust \cite{Mathis1977}, and our choice has been commonly adopted in multi-wavelength radiative transfer modeling of protoplanetary disks \cite{Wolf2003,liuy2012}. For the LGP, we set $a_{\rm{min}}\,{=}\,0.1\,\mu{\rm m}$ and $a_{\rm{max}}\,{=}\,1\,{\rm mm}$. Finally, we derived the averaged dust opacities using the mass fractions of each dust species for the IRS spectrum presented in Table~\ref{tab:comp}. 

\subsubsection{Fit to the IRS and MIRI spectra}
\label{sec:fitirs}

Our modeling strategy is to first search for a parameter set that reproduces the IRS spectrum. Then, we changed the inner disk structure and dust density distribution to fit the MIRI spectrum. As described in the above section, there are five free parameters ($H_{\rm 100.SGP}$, $\beta$, $p$, $\eta$, $M_{\rm dust}$) that are expected to have significant effects on the infrared SED. To reduce the dimensionality of the parameter space, we fixed the total dust mass to the value that is converted from measured millimeter flux densities. Assuming the emitting dust is optically thin, the dust mass can be analytically calculated via $M_{\rm dust}\,{=}\,F_{\nu}d^2/\left(\kappa_{\nu}B_{\nu}({T_{\rm dust}^{\rm mean}})\right)$, where ${T_{\rm dust}^{\rm mean}}$ is the mean dust temperature. Literature studies either assumed a constant ${T_{\rm dust}^{\rm mean}}\,{=}\,20\,{\rm K}$ \cite{Manara2023} or adopted a value that is dependent on stellar luminosity \cite{Andrews2013,Liu2024}. For simplicity, we took $T_{\rm dust}^{\rm mean}\,{=}\,20\,{\rm K}$. We also note that the assumed mean dust temperature is only used to set up the dust mass. In the radiative transfer models, the temperature structure of the disk is self-consistently derived by \texttt{RADMC-3D}. At the frequency of $\nu\,{=}\,225\,\rm{GHz}$ (or $\lambda\,{=}\,1.33\,{\rm mm}$), the measured flux density is $F_{\nu}\,{=}\,57.6\,{\rm mJy}$ \cite{Facchini2021}, and the dust absorption opacity is $\kappa_{\nu}\,{=}\,1.9\,{\rm cm^2/g}$ for the LGP, yielding $M_{\rm dust}\,{=}\,8\,{\times}\,10^{-5}\,M_{\odot}$. After a detailed parameter study, we found that a combination of $H_{\rm 100.SGP}\,{=}\,8\,\rm{AU}$, $\beta\,{=}\,1.1$, $p\,{=}\,0.3$, and $\eta\,{=}\,3\,{\times}\,10^{-3}$ provides a good match with the IRS spectrum. This model is referred as model\,A, with $\Sigma_{\rm dust}$ indicated with the grey line in the left panel of Figure~\ref{fig:surdens}. A comparison between the model spectrum and IRS data is shown in the middle panel of Figure~\ref{fig:surdens}, whereas the right panel demonstrates that model\,A also reproduces the photometric data points from optical to millimeter wavelengths. Model parameters are tabulated in Table~\ref{tab:paras}. The scale height and flaring index of model\,A are quite typical for protoplanetary disks. The shallow profile of the dust surface density with $p\,{=}\,0.3$ distributes small dust grains toward the cool outer part of the inner disk, and hence produces strong emission at $\lambda\,{\gtrsim}\,16\,\mu\mathrm{m}$ that is consistent with the observed IRS level.

Starting from model\,A, we varied the parameters of the inner disk to investigate how the infrared SED is influenced. The properties of the outer disk were all maintained in the experiment (see Table~\ref{tab:paras}), because their impacts on the SED are predominately in the far-infrared and millimeter. Moreover, we also fixed the flaring index $\beta\,{=}\,1.1$ to reduce the model degeneracy. To fit the MIRI spectrum, we found that small dust grains in the inner disk need to be further depleted with $\eta\,{=}\,5\,{\times}\,10^{-4}$, and are more concentrated toward the hot inner region with a surface density power-law index of $p\,{=}\,0.7$. In addition, the disk scale height of the inner disk needs to be increased from 8\,AU to 12\,AU to bring the ${\sim}\,10\,\mu{\rm m}$ silicate feature up to the observed MIRI level. In Figure~\ref{fig:surdens}, the black line in the left panel shows $\Sigma_{\rm dust}$ of the new model, i.e., model\,B, whereas the corresponding infrared spectrum and broad-band SED are displayed in the middle and right panels, respectively. Our modeling results suggest that the inner disk of PDS\,70 is dynamic. However, we note that the adjustment of the inner disk structure is not a peculiarity of PDS\,70, but it is also observed in other protoplanetary disks \cite{Flaherty2012,Fang2014,Flaherty2016}. 

Comparing model\,B with model\,A, the inner disk lost a dust mass of ${\sim}\,10^{-8}\,M_{\odot}$ over 15 years. The total dissipated mass of dust and gas is then $3\times10^{-7}\,M_{\odot}$ assuming a gas-to-dust mass ratio of 30 \cite{Ansdell2016,Ansdell2017,Miotello2023}. This amount of material cannot be explained by the accretion process alone, because the accretion rate of PDS\,70 is quite low, approximately $10^{-10}\,M_{\odot}/{\rm yr}$ \cite{Manara2019,Skinner2022,Gaidos2024}. Photoevaporation and disk winds can contribute to the mass loss \cite{Picogna2019,Pascucci2023,Fang2023}. 

\begin{figure}[H]
\centering \includegraphics[width=0.45\textwidth,angle=0]{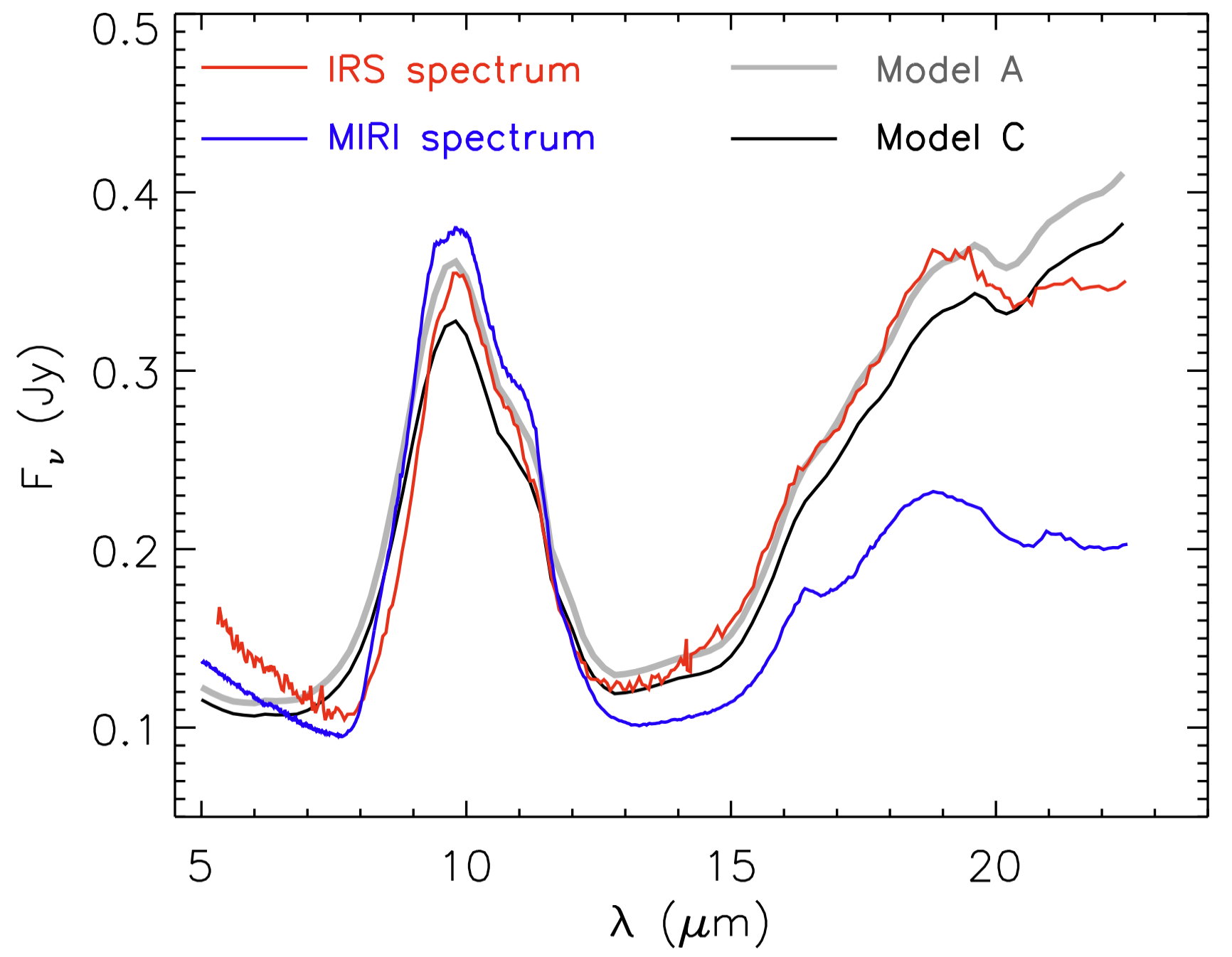}
\caption{Comparison of radiative transfer models with different grain sizes for the SGP. Parameter values of model\,A are given in Table~\ref{tab:paras}. Note that model\,A and model\,C differ only in terms of grain sizes for the SGP, see Sect.~\ref{sec:graingrowth}.}
\label{fig:graingrowth}
\end{figure}

\subsubsection{Pure grain growth as an explanation of the change in mid-infrared fluxes?}
\label{sec:graingrowth}
As demonstrated in Sect.~\ref{sec:specdecom}, an increase in mass-averaged grain size is required to reproduce the variation of distinctive silicate features. We also examined whether or not grain growth alone can interpret the overall change in mid-infrared flux densities. In model\,A, the minimum and maximum grain size were set to $a_{\rm min}\,{=}\,0.005\,\mu{\rm m}$ and $a_{\rm max}\,{=}\,0.1\,\mu{\rm m}$ for the SGP, respectively. We set $a_{\rm min}\,{=}\,0.01\,\mu{\rm m}$ and $a_{\rm max}\,{=}\,0.2\,\mu{\rm m}$, both of which are increased by a factor of two. By keeping all of other parameters identical to those of model\,A, we calculated a new model referred as model\,C.

The results are shown in Figure~\ref{fig:graingrowth}. It is clear to see that pure grain growth leads to an overall reduction of flux from the near- to mid-infrared wavelength regime, which contradicts the observation. Dust grains with larger sizes have lower opacities at optical and near-infrared wavelengths, which in turn results in lower dust temperature from self-consistent radiative transfer simulation. Consequently, the model exhibits an overall decrease in infrared fluxes.

\subsection{Other possibilities}

We have explored the changes in scale height and dust density distribution of the inner disk as the interpretation of the variability in mid-infrared fluxes. There may be other possibilities. An inner rim, if present, can cast a shadow across the outer disk. Increasing the rim height or size can cause a reduction of flux in the mid-infrared regime. However, the puffed-up inner rim would produce more near-infrared flux that is not consistent with the observation. High-resolution images reveal that some protoplanetary disks have warps or misaligned inner disk regions \cite{Perez2018,Bi2020}. These out-of-plane substructures can also cast shadows, and obscure stellar photons to cause variability \cite{Marino2015,Benisty2018,Ansdell2020}. For PDS\,70, warps may be generated due to the gravitational perturbation by the third protoplanet candidate embedded within the inner disk. Implementing these scenarios needs three-dimensional setups, and requires abundant inputs, e.g., the radial and azimuthal extent of the warp, and the misalignment angle of the inner disk region \cite{Whitney2013,Liu2022}. Given the fact that constraints on these properties are not available from current observations, we leave it as a future direction. 
   
\section{Summary}
\label{sec:sum}

We analyzed the IRS and MIRI spectra of the PDS\,70 disk observed at two different epochs, with the goal of investigating the dust composition, dust size, and their variation over time. The results show that both the mass-averaged dust size and crystallinity increase by a factor of about two over a period of 15 years, indicating evident dust processing in the terrestrial planet-forming region. Moreover, the dust size and crystallinity in the PDS\,70 disk are found to be similar to those of other disks, which implies that the two giant planets, PDS\,70b and PDS\,70c located in the prominent gap, do not have a significant effect on the dust processing in the innermost zone of the disk.

A comparison between the IRS and MIRI spectra shows that the PDS\,70 disk is variable in the mid-infrared. Through self-consistent radiative transfer modeling, we investigated the cause of such a variability. Our modeling efforts suggest that the inner disk of PDS\,70 is dynamic. Adjustments to the scale height and dust density distribution in the inner disk are able to produce a bit stronger ${\sim}\,10\,\mu{\rm m}$ silicate feature and obviously reduce the long-wavelength ($\lambda \,{\sim}\, 16\,\mu \mathrm {m}$) flux in the meanwhile, consistent with the observed trend from the IRS to MIRI observations. Future high-resolution and high-sensitivity observations are needed to validate the hypothesis.

\Acknowledgements{YL acknowledges financial supports by the National Natural Science Foundation of China (Grant number 11973090), and by the International Partnership Program of Chinese Academy of Sciences (Grant number 019GJHZ2023016FN). HW acknowledges the financial support by the National Natural Science Foundation of China (Grant number 11973091). FD acknowledges financial supports by the National Natural Science Foundation of China with grant 12041305 and the National Key R\&D Program of China with grant 2023YFA1608000. GP gratefully acknowledges support from the Carlsberg Foundation, grant CF23-0481 and from the Max Planck Society. In this study, a cluster is used with the SIMT accelerator made in China. The cluster includes many nodes each containing 4 CPUs and 8 accelerators. The accelerator adopts a GPU-like architecture consisting of a 16GB HBM2 device memory and many compute units. Accelerators connected to CPUs with PCI-E, the peak bandwidth of the data transcription between main memory and device memory is 16GB/s. This work is partially supported by cosmology simulation database (CSD) in the National Basic Science Data Center (NBSDC-DB-10).}


\InterestConflict{The authors declare that they have no conflict of interest.}



\bibliographystyle{scichina}
\bibliography{pds70}

\begin{appendix}

\renewcommand{\thesection}{Appendix}

\section{Radiative transfer models with different dust opacities}
\label{sec:moredust}

\begin{figure*}[!t]
 \centering
 \includegraphics[width=0.9\textwidth]{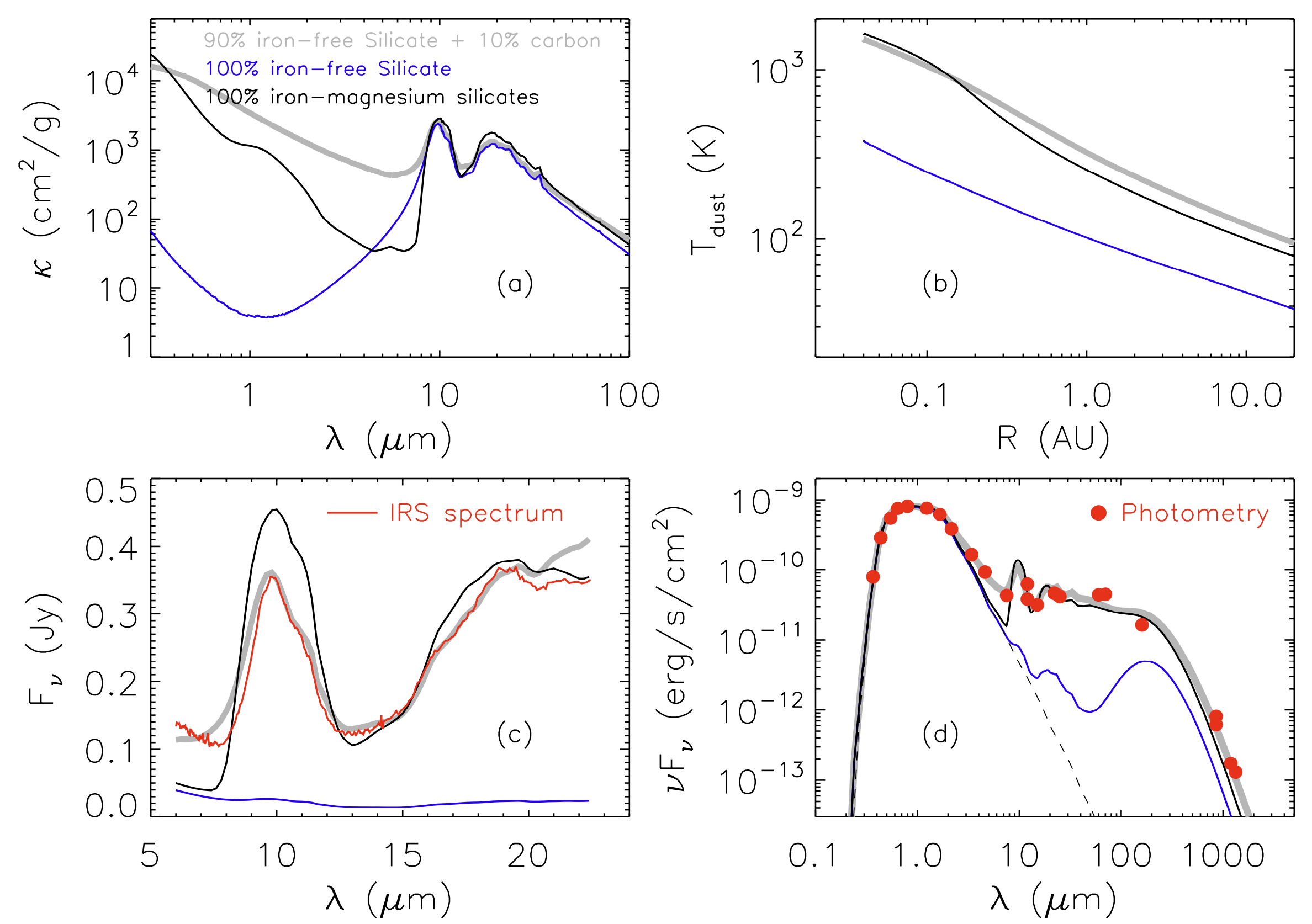}
 \caption{Comparison of radiative transfer models with different types of dust grains. The grey lines show the results for iron-free silicate mixed with carbon. The results for 
 pure iron-free silicate are indicated with blue lines. The black lines show the results for iron-magnesium silicates. {\it Panel (a):} mass absorption coefficients of different types of dust. {\it Panel (b):} dust temperature in the surface layer of the disk obtained from the radiative transfer modeling with \texttt{RADMC3D}. {\it Panel (c):} model infrared spectra compared to the IRS spectrum. {\it Panel (d):} model SEDs compared to the photometric data points from the optical to millimeter wavelengths. Note that all the parameters of models in comparison are identical, and are taken from model A, see Table~\ref{tab:paras}.}
\label{fig:moredust}
\end{figure*}

In Sect.~\ref{sec:rtmodel}, we conducted a detailed radiative transfer modeling of PDS\,70 in order to investigate the mid-infrared variability. The dust ensemble was assumed to be a mixture of silicate grains and carbon grains, with the volume fraction of carbon being 10\%. Carbon grains were incorporated to increase the mass absorption coefficients at optical and near-infrared wavelengths, which are necessary to obtain relatively high temperature from self-consistent radiative transfer simulations. Panel (a) in Figure~\ref{fig:moredust} shows a comparison of mass absorption coefficients ($\kappa$) between different types of dust grains. As can be seen, pure iron-free silicate dust features a much lower $\kappa$ than it mixed with carbon. Such a weak capability of absorbing stellar energy leads to much lower dust temperature ($T_{\rm dust}$) in the disk surface layer, see panel (b). As a consequence, the silicate emission feature is very weak, and the predicted fluxes from the infrared all the way to millimeter wavelength regime are quite low, see panel (c) and (d). 

Moreover, we also checked the results when considering iron-magnesium silicates (${\rm MgFeSiO_4}$ and ${\rm MgFeSi_2O_6}$, \cite{Dorschner1995}) in the radiative transfer models. 
Generally, mass absorption coefficients of iron-magnesium silicates lie between those of iron-free silicate and iron-free silicate mixed with carbon. However, at $\lambda\,{\lesssim}\,0.4\,\mu{\rm m}$, iron-magnesium silicates have the largest $\kappa$ among the three types of dust in comparison. Consequently, the innermost disk region ($R\,{\lesssim}\,0.1\,{\rm AU}$) is the hottest, and beyond that radial location ($R\,{\gtrsim}\,0.1\,{\rm AU}$) the dust gets an intermediate temperature. As can be seen in 
panel (a) of Figure~\ref{fig:moredust}, the contrast of the $10\,\mu{\rm m}$  feature in the $\kappa$ curve is the highest for iron-magnesium silicates, which in turn results in the strongest silicate emission feature, see panel (c). In the meanwhile, model fluxes at $\lambda\,{\lesssim}\,8\,\mu{\rm m}$ are obviously lower than the observed level. Compared to the case of iron-free silicate mixed with carbon, the outer disk is cooler, leading to lower fluxes at (sub-)millimeter regimes.

\end{appendix}

\end{multicols}
\end{document}